\documentclass[12pt]{article}
\usepackage{graphicx}

\input{epsf}
\usepackage{cite}
\def\@fmsl@sh#1#2#3{\m@th\ooalign{$\hfil#1\mkern#2/\hfil$\crcr$#1#3$}}
 \def\eq#1\en{\begin{equation}#1\end{equation}}
\def\s[#1,#2]{[#1\stackrel{\star}{,}#2]}
\def\sx[#1,#2]{[#1\stackrel{\star_{x}}{,}#2]}

\newcommand{\nc}{\newcommand}
\nc{\beq}{\begin{equation}}
\nc{\eeq}{\end{equation}}
\nc{\beqa}{\begin{eqnarray}}
\nc{\eeqa}{\end{eqnarray}}

\def\a {\alpha}

\def\gsim{\mathrel{\rlap{\lower4pt\hbox{\hskip1pt$\sim$}}
    \raise1pt\hbox{$>$}}}       

\begin{document}
\makeatletter
\def\fmslash{\@ifnextchar[{\fmsl@sh}{\fmsl@sh[0mu]}}
\def\fmsl@sh[#1]#2{%
  \mathchoice
    {\@fmsl@sh\displaystyle{#1}{#2}}%
    {\@fmsl@sh\textstyle{#1}{#2}}%
    {\@fmsl@sh\scriptstyle{#1}{#2}}%
    {\@fmsl@sh\scriptscriptstyle{#1}{#2}}}
\def\@fmsl@sh#1#2#3{\m@th\ooalign{$\hfil#1\mkern#2/\hfil$\crcr$#1#3$}}
\makeatother


\title{\large{\bf Quantum Electrodynamics on Noncommutative Spacetime}}

\author{Xavier~Calmet\thanks{xcalmet@ulb.ac.be} \\
Service de Physique Th\'eorique, CP225 \\
Boulevard du Triomphe \\
B-1050 Brussels \\
Belgium 
}

\date{June, 2006}

\maketitle

\begin{abstract}
We propose a new method to quantize gauge theories formulated on a
canonical noncommutative spacetime with fields and gauge
transformations taken in the enveloping algebra. We show that the
theory is renormalizable at one loop and compute the beta function and
show that the spin dependent contribution to the anomalous magnetic moment of the 
fermion at one loop has the same value as in the commutative quantum electrodynamics case.

\end{abstract}


\newpage


Gauge theories formulated on a canonical noncommutative spacetime have
recently received a lot of attention, see e.g. \cite{Douglas:2001ba} for a review. A canonical noncommutative
spacetime is defined by the noncommutative algebra
\begin{eqnarray} \label{NCA}
[ \hat x^\mu,\hat x^\nu]=i\theta^{\mu\nu}
\end{eqnarray}
where $\mu$ and $\nu$ run from 0 to 3 and where $\theta^{\mu\nu}$ is
constant and antisymmetric. It has mass dimension minus
two. Formulating Yang-Mills theories relevant to particle physics on
such a spacetime requires to consider matter fields, gauge fields and
gauge transformations in the enveloping algebra otherwise SU(N) gauge
symmetries cannot be implemented. It has been pointed out that actions
formulated on a canonical noncommutative spacetime are invariant under
noncommutative Lorentz transformations \cite{Calmet:2004ii} which
preserve the algebra (\ref{NCA}) and thus the minimal length implied
by the relation (\ref{NCA}).

The enveloping algebra approach \cite{Madore:2000en,Jurco:2000ja,Jurco:2001rq,Calmet:2001na} allows to map a noncommutative action
$\hat S$ on an effective action formulated on a regular commutative
spacetime. The dimension four operators are the usual ones and the
noncommutative nature of spacetime is encoded into  higher order
operators. In this paper we will be considering quantum electrodynamics on a noncommutative spacetime. The noncommutative action for a Dirac fermion
coupled to a U(1) gauge field is given by
\begin{eqnarray} \label{ncqed}
\int d^4x \bar{\hat \Psi}(\hat x) (i\hat{\fmslash D} -m) \hat \Psi(\hat x) -\frac{1}{4} \int d^4x \hat F_{\mu\nu} (\hat x) 
\hat F^{\mu\nu} (\hat x)
\end{eqnarray}
where the hat on the coordinate $x$ indicates that the functions belong
to the algebra of noncommutative functions and the hat over the
functions that they are to be considered in the enveloping algebra. 
The procedure \cite{Madore:2000en,Jurco:2000ja,Jurco:2001rq,Calmet:2001na}
to map actions such as (\ref{ncqed}) on an effective
action formulated on a commutative spacetime requires one to first define
a vector space isomorphism that maps the algebra of noncommutative
functions on the algebra of commutative functions. The price to pay for
replacing the noncommutative argument of the function by a commutative
one is the introduction of a star product: $f(\hat x) g(\hat x) = f(x)
\star g(x)$.  One then expands the fields in the enveloping algebra
using the Seiberg-Witten maps \cite{Seiberg:1999vs} and obtains:
\begin{eqnarray}
\label{actionqed}
\int  \bar{\hat\Psi}(\hat x)  (i \fmslash{\hat D}-m) \hat \Psi (\hat x) d^4x 
 &=&
 \int  \bar{\psi} (i \fmslash{D}- m)\psi d^4x  \\ 
 \nonumber 
 && -\frac{1}{4} g \int 
\theta^{\mu \nu}
 \bar{\psi} F_{\mu \nu} (i \fmslash{D} -m )\psi d^4x 
-\frac{1}{2} g  \int \theta^{\mu \nu}
 \bar{\psi} \gamma^\rho F_{\rho \mu} i D_\nu \psi  d^4x 
\\  
-\frac{1}{4} \int \hat F_{\mu \nu } (\hat x)
 \hat F^{\mu \nu}(\hat x) d^4 x&=&-\frac{1}{4} \int  F_{\mu \nu }
 F^{\mu \nu } d^4x
 \\ \nonumber &&
 +\frac{1}{8}  g
\int \theta^{\sigma \rho } F_{\sigma \rho }F_{\mu \nu } F^{\mu \nu } d^4x 
-\frac{1}{2}g  \int  \theta^{\sigma \rho } F_{\mu \sigma}F_{\nu \rho} F^{\mu \nu} d^4x,
\end{eqnarray} 
to  first order in $\theta^{\mu\nu}$ and where as usual $F^{\mu\nu}=\partial^\mu A^\nu - \partial^\nu A^\mu$.

The standard procedure to quantize this action is to use the expanded
version (\ref{actionqed}) and decompose the classical degrees of
freedom i.e. $\Psi$ and $A_\mu$ in terms of creation and annihilation
operators, see e.g. \cite{Carlson:2001sw, Martin:2002nr}. One then
finds that the theory is not renormalizable
\cite{Wulkenhaar:2001sq}. We propose a new approach to the
quantization of noncommutative gauge theories.  It should be stressed that this new approach is fundamentally different from the one traditionally followed. Our approach allows to perform
consistent quantum calculations at least at the one loop level.

We start as usual from the action (\ref{ncqed}) which is invariant under U(1) gauge transformations and use the vector space isomorphism to replace the noncommutative arguments of the functions by commutative ones:
\begin{eqnarray}
\label{iso}
\int  \bar{\hat\Psi}(\hat x)  (i \fmslash{\hat D}-m) \hat \Psi (\hat x) d^4x 
 &=&
\int  \bar{\hat\Psi}(x) \star (i \fmslash{\hat D}-m) \hat \Psi (x) d^4x
\\  \nonumber
-\frac{1}{4} \int \hat F_{\mu \nu } (\hat x)
 \hat F^{\mu \nu}(\hat x) d^4 x&=&-\frac{1}{4} \int \hat F_{\mu \nu } (x) \star
 \hat F^{\mu \nu}(x) d^4 x
 \end{eqnarray} 
i.e. we have not yet expanded the fields in the enveloping algebra. We
then quantize the action (\ref{iso}) using the fields in the
enveloping algebra $\hat \Psi(x)$ and $\hat A^\mu(x)$ which we
decompose in terms of creation $\hat a^\dagger$ and annihilation operators
$\hat a$ which are themselves in the enveloping algebra. We impose
the usual algebra for the hatted creation and annihilation operators:
 \begin{eqnarray} \label{quant}
 [\hat a,\hat a^\dagger]_{\mp}=1, \ [\hat a,\hat a]_{\mp}=0 \
 \mbox{and} \ [\hat a^\dagger,\hat a^\dagger]_{\mp}=0,
 \end{eqnarray} 
 where as usual the minus sign refers to a commutator for bosons and
the plus sign to an anticommutator for fermions. Note that $\hat \Psi( x)$ transforms as a spinor under noncommutative Lorentz transformations \cite{Calmet:2004ii}  and $\hat A^\mu(x)$ as a vector under the same transformations. It thus makes sense to impose the relations (\ref{quant}).   Furthermore asymptotically the fields $\hat \Psi( x)$ and $\Psi( x)$ are equivalent since the expansion in the enveloping algebra is an expansion in $g \theta^{\mu\nu}$, one thus finds that in the limit $g\to0$ one recovers the usual creation and annihilation operators as it should be for the physical states. We now procedure further with the quantization of the action (\ref{actionqed}).
As in the Lie algebra case \cite{Martin:1999aq,Hayakawa:1999yt,Hayakawa:1999zf}  we
use the BRST quantization procedure and introduce the gauge fixing and
Faddeev-Popov terms
\begin{equation}
 S_{\rm GF} =
  \int d^4 x
  \left(
   -\frac{1}{2\alpha} \partial_\mu \hat A^\mu \star \partial_\nu \hat A^\nu
   +
   \frac{1}{2}
   \left(
    i\bar{\hat c} \star \partial^\mu \hat D_\mu \hat c -
    i\partial^\mu \hat D_\mu \hat c \star \bar{\hat c}
   \right)
  \right).
\end{equation}
We thus recover the Feynman rules given in \cite{Martin:1999aq,Hayakawa:1999yt,Hayakawa:1999zf}  for the enveloping algebra valued fields which are not yet the physical degrees of freedom on the commutative spacetime but which can be used to quantize the theory. The Feynman rules are:
\begin{eqnarray}
 \hat \Psi(p^I) \hat A^\mu \hat \Psi(p^F) &\to& i g \gamma^\mu \exp(
 \frac{i}{2} p_\alpha^I \theta^{\alpha\beta} p^F_\beta) \\ \hat
 A^{\mu_1}(p^1) \hat A^{\mu_2}(p^2) \hat A^{\mu_3}(p^3) &\to& -2 g \sin(
 \frac{1}{2} p_\mu^1 \theta^{\mu\nu} p^2_\nu) \\ && \nonumber
 [(p^1-p^2)^{\mu_3} g^{\mu_1\mu_2} + (p^2-p^3)^{\mu_1} g^{\mu_2\mu_3}
 +(p^3-p^1)^{\mu_2} g^{\mu_3\mu_1}] \\ \hat A^{\mu_1}(p^1) \hat
 A^{\mu_2}(p^2) \hat A^{\mu_3}(p^3) \hat A^{\mu_4}(p^4) &\to& - 4 i g^2 [ (
 g^{\mu_1\mu_3}g^{\mu_2\mu_4} - g^{\mu_1\mu_4}g^{\mu_2\mu_3}) \\ &&
 \nonumber \sin( \frac{1}{2} p_\mu^1 \theta^{\mu\nu} p^2_\nu) \sin(
 \frac{1}{2} p_\alpha^3 \theta^{\alpha\beta} p^4_\beta) \\ &&
 \nonumber + ( g^{\mu_1\mu_4}g^{\mu_2\mu_3} -
 g^{\mu_1\mu_2}g^{\mu_3\mu_4}) \sin( \frac{1}{2} p_\mu^3
 \theta^{\mu\nu} p^1_\nu) \sin( \frac{1}{2} p_\alpha^2
 \theta^{\alpha\beta} p^4_\beta) \\ \nonumber && + (
 g^{\mu_1\mu_2}g^{\mu_3\mu_4} - g^{\mu_1\mu_3}g^{\mu_2\mu_4}) \sin(
 \frac{1}{2} p_\mu^1 \theta^{\mu\nu} p^4_\nu) \sin( \frac{1}{2}
 p_\alpha^2 \theta^{\alpha\beta} p^3_\beta)] \\ \hat c(p^I) \hat A^\mu
 \hat c(p^F) &\to& 2 i g p^F_\mu \sin( \frac{1}{2} p_\alpha^I
 \theta^{\alpha\beta} p^F_\beta).
\end{eqnarray}
It turns out that the theory at least at one loop is renormalizable \cite{Hayakawa:1999yt,Hayakawa:1999zf}. 
Using the results of \cite{Hayakawa:1999yt,Hayakawa:1999zf} it is straightforward to obtain the 
$\beta$ function which is given by
\begin{equation}
 \beta(g) = \frac{1}{g} Q \frac{dg}{dQ}
  = -\left(
       \frac{22}{3} - \frac{4}{3} 
     \right)\, \frac{g^2}{16\pi^2},
   \label{beta}
\end{equation}
 where the contribution $22/3$ is due to the structure of the gauge interaction which is similar
to that of the nonabelian SU(2) Yang-Mills theory \cite{Martin:2002nr} and the factor $4/3$ \cite{Hayakawa:1999yt,Hayakawa:1999zf} is due to the fermion field $\hat \psi$.  Using the results of \cite{Hayakawa:1999yt,Hayakawa:1999zf} and \cite{Riad:2000vy} it is easy to obtain the renormalized vertex for the fermion-gauge field vertex $\bar{\hat \Psi} \star \hat A_\mu \Gamma^\mu \star \hat \Psi$: 
\begin{eqnarray} \label{renvertex}
\Gamma^\mu= E_1 \gamma^\mu + H_1 (p'+p)^\mu- G_1 \theta^{\mu\nu} q_\nu - E_2 \gamma^\mu p_\alpha  \theta^{\alpha\beta} q_\beta - H_3 (p'+p)^\mu \gamma_\alpha \theta^{\alpha\beta} q_\beta
\end{eqnarray} 
where $q^\mu=(p'-p)^\mu$ and the functions $E_1$, $H_1$, $G_1$, $E_2$ and $H_3$ were calculated in \cite{Riad:2000vy} , we give them in the appendix.

One can then perform the second map and expand the fields in the
enveloping algebra using the Seiberg-Witten maps which are now at the
quantum level and obtain the action to first order in $\theta^{\mu\nu}$
\begin{eqnarray}
\label{actionqedrn}
\int \bar{\hat\Psi}(\hat x) (i \fmslash{\hat D}-m) \hat \Psi (\hat x)
 d^4x &=& \int \bar{\psi} (i \fmslash{D}- m)\psi d^4x \\ \nonumber &&
 -\frac{1}{4} g \int \theta^{\mu \nu} \bar{\psi} F_{\mu \nu} (i
 \fmslash{D} -m )\psi d^4x -\frac{1}{2} g \int \theta^{\mu \nu}
 \bar{\psi} \gamma^\rho F_{\rho \mu} i D_\nu \psi d^4x \\ -\frac{1}{4}
 \int \hat F_{\mu \nu } (\hat x) \hat F^{\mu \nu}(\hat x) d^4
 x&=&-\frac{1}{4} \int F_{\mu \nu } F^{\mu \nu } d^4x \\ \nonumber &&
 +\frac{1}{8} g \int \theta^{\sigma \rho } F_{\sigma \rho }F_{\mu \nu
 } F^{\mu \nu } d^4x -\frac{1}{2}g \int \theta^{\sigma \rho } F_{\mu
 \sigma}F_{\nu \rho} F^{\mu \nu} d^4x,
\end{eqnarray} 
where the coupling constant, the mass and the fields are now renormalized. The renormalized fields $\psi$ and $A^\mu$ are the physical degrees of freedom. The renormalized vertex $\bar \psi \Gamma_\mu A^\mu \psi$ is given by:
\begin{eqnarray} \label{renvertex2}
\Gamma^\mu= E_1 \gamma^\mu + H_1 (p'+p)^\mu- G_1 \theta^{\mu\nu} q_\nu - E_2 \gamma^\mu p_\alpha  \theta^{\alpha\beta} q_\beta - H_3 (p'+p)^\mu \gamma_\alpha \theta^{\alpha\beta} q_\beta.
\end{eqnarray} 
The functions $E_1$, $H_1$, $G_1$, $E_2$ and $H_3$ are given in the appendix. It is easy to see that the form factors $F^1(0)$ and $F^2(0)$ have the usual QED values.

This approach to the quantization of noncommutative gauge theories
allows one to make consistent quantum calculations at least at the one
loop level in the enveloping algebra approach. It implies that only
the operators generated via the Seiberg-Witten maps are compatible
with the noncommutative gauge invariance and that further operators
discussed in the literature were artifacts of the quantization and
regularization procedures. The bounds on spacetime noncommutativity
are thus only of the order of a few TeV \cite{Calmet:2004dn} and it
remains important to try to improve these bounds.

Physical observables can be calculated at the quantum level
independently of a cutoff. For example one finds using the result of \cite{Hayakawa:1999yt,Hayakawa:1999zf,Riad:2000vy} that the anomalous
magnetic moment of a fermion has the usual quantum electrodynamics
value. More precisely the noncommutative contribution is spin independent and will thus not contribute to the anomalous magnetic moment  \cite{Kersting:2001zz} when measured with classical methods. 
Once we expand the action in $\theta$, we see that the leading order contribution to the magnetic moment of the fermion is given by
\begin{eqnarray}
\frac{e}{m} \left [ \left ( F^1(0) + F^2(0) \right) \vec S + \frac{\alpha  \gamma_{\mbox{{\tiny Euler}}}}{6 \pi} m^2 \vec \theta \right ],
\end{eqnarray} 
i.e. the spin dependent part of the magnetic moment is the usual QED value. One finds also a contribution to the electric dipole moment
\begin{eqnarray}
\frac{1}{4} e (\vec \theta \times \vec p) \left ( 1 + \frac{3}{\pi} \alpha \gamma_{\mbox{{\tiny Euler}}} \right)
\end{eqnarray} 
which is purely a noncommutative effect. The form factors are the usual ones and are defined in the appendix. 

The value of the $\beta$-function is an issue from a phenomenological point of view as
it does not match that of quantum electrodynamics on a commutative
spacetime.  However,  it is probable that the noncommutative parameter $\theta^{\mu\nu}$ is not
a simple constant but is spacetime or energy dependent as studied in
\cite{Calmet:2003jv}. The assumption that $\theta^{\mu\nu}$ is scale dependent is not that far-fetched, indeed if it is the expectation value of a background field, as e.g. in the string theory picture \cite{Seiberg:1999vs}, one would expect a scale dependence of the renormalized expectation value. 
It is also clear that the UV/IR mixing
\cite{Minwalla:1999px} will also appear in the enveloping algebra
approach and although this approach allows to implement any gauge
symmetry, it does not improve the issues linked to the energy behavior of
the theory. Furthermore  UV/IR mixing is expected to jeopardize the renormalizability of the theory at higher loops, but this remains to be verified. 
Finally we want to stress that although we have applied
our approach to a noncommutative U(1) theory only, it can be easily
extended to any gauge group. We emphasize that our approach is fundamentally different from that developed in e.g. \cite{Carlson:2001sw,Martin:2002nr,Wulkenhaar:2001sq}. Our method  is fully in the spirit of the enveloping algebra approach where the full theory, in our case the quantized theory, is mapped on a commutative spacetime to identify the physical degrees of freedom.  One of the main differences is that when quantized our way, the action is renormalizable, which is not the case in  \cite{Carlson:2001sw,Martin:2002nr,Wulkenhaar:2001sq} where any observable is cutoff dependent.

\section*{Appendix}
In this appendix we reproduce the functions $E_1$, $H_1$, $G_1$, $E_2$ and $H_3$ which were calculated in \cite{Riad:2000vy}:
\begin{eqnarray}
E_1&=& \frac{-\alpha}{\pi} \int^1_0 d\alpha_1 d\alpha_2 d\alpha_3 \delta\left(1 -\sum \alpha_i \right) \times \left ( 1 - e^{i (\alpha_2+\alpha_3)  p \cdot  \tilde q} e^{-i  p \times p'}\right) \times \nonumber \\ \nonumber
&& \times \left \{  \left [ \frac{(2 p'\cdot p-(\alpha_2+\alpha_3)(p'+p)^2)+m^2 (\alpha_2+\alpha_3)^2 - \alpha_2 \alpha_3 q^2}{2(\alpha_1 m_\gamma^2 +(\alpha_2+\alpha_3)^2 m^2 -\alpha_2 \alpha_3 q^2) }+ \gamma_{\mbox{{\tiny Euler}}}\right] e^{-i (\alpha_2+\alpha_3) p\cdot \tilde q}  \right. \\ 
&& \left. +\left[ \frac{(\alpha_2+\alpha_3)(p'+p)^2 - 3 m^2 - m^2 (\alpha_2 +\alpha_3)^2 +\alpha_2 \alpha_3 q^2}{m^2( \alpha_1-\alpha_2 -\alpha_3)+ m_\gamma^2 (\alpha_2 + \alpha_3) + m^2 (\alpha_2 +\alpha_3)^2 - \alpha_2 \alpha_3 q^2} - \frac{3 \gamma_{\mbox{{\tiny Euler}}}}{2} \right] \right \},
\end{eqnarray} 

\begin{eqnarray}
H_1&=& \frac{-\alpha}{\pi}
\int^1_0 d\alpha_1 d\alpha_2 d\alpha_3 \delta\left(1 -\sum \alpha_i \right) \times \nonumber \\ 
&&
\times \left \{  \frac{m \a_1 (\a_2+\a_3) e^{i(\a_2+\a_3)p\cdot\tilde q}}{\a_1 m^2_\gamma + (\a_2+\a_3)^2 m^2 -\a_2\a_3q^2} + \right.  \\ && \nonumber
+ \left.  \frac{m \a_1 (\a_2+\a_3) \left( 1 - e^{i(\a_2+\a_3) p\cdot \tilde q} e^{-i p\times p'} \right) }{m^2 (\a_1-\a_2-\a_3) + m_\gamma^2 (\a_2+\a_3) + m^2 (\a_2+\a_3)^2 -\a_2 \a_3 q^2} \right \},
\end{eqnarray}

\begin{eqnarray}
G_1&=& \frac{-\alpha}{\pi} i m \gamma_{\mbox{{\tiny Euler}}}
\int^1_0 d\alpha_1 d\alpha_2 d\alpha_3 \delta\left(1 -\sum \alpha_i \right) \times \nonumber \\ 
&&
\times 
\left ( ( 1+\alpha_2 + \alpha_3) e^{-i (\alpha_2 + \a_3) p \cdot \tilde q} - ( 2 -\a_2 -\a_3) e^{i(\a_2 + \a_3) p \cdot \tilde q} e^{- i p \times p'} \right),
\end{eqnarray} 

\begin{eqnarray}
E_2&=& \frac{-\alpha}{\pi} i  \gamma_{\mbox{{\tiny Euler}}}
\int^1_0 d\alpha_1 d\alpha_2 d\alpha_3 \delta\left(1 -\sum \alpha_i \right) \times \nonumber \\ 
&&
\times 
\left ( 1- e^{-i p \times p'} \right) ( 2 -\alpha_2 -\alpha_3) e^{-i(\alpha_2+\a_3) p\cdot \tilde q},
\end{eqnarray} 

\begin{eqnarray}
H_3&=& \frac{-\alpha}{\pi} \frac{i  \gamma_{\mbox{{\tiny Euler}}}}{2}
\int^1_0 d\alpha_1 d\alpha_2 d\alpha_3 \delta\left(1 -\sum \alpha_i \right) \times \nonumber \\ 
&&
\times 
\left ( (2 -\a_2-\a_3) e^{-i (\a_2+ \a_3) p \cdot \tilde q} + (1 + \a_2 + \a_3) e^{i(\a_2+\a_3) p\cdot \tilde q} 
e^{-i p\times p'} \right),
\end{eqnarray} 
where we have used the following notation  $p\cdot \tilde q= p_\mu \theta^{\nu\mu} q_\nu$, $p\times p'= p_\mu \theta^{\mu\nu} p'_\nu$, $\tilde q^\mu =\theta^{\nu\mu} q_\nu$ and where $m_\gamma$ is a small photon mass that takes care of the IR divergences. Note that the exponential corresponding to the star product is not included into these functions since we calculate the renormalized vertex $\bar{\hat \Psi} \star \Gamma_\mu  \hat A^\mu \star \hat \Psi$ and we still have to expand the star product and the fields in the enveloping algebra to obtain the physical degrees of freedom.

As shown in \cite{Riad:2000vy}, the leading order non-relativistic limits of these functions are given by:
\begin{eqnarray}
F_1(q^2)&=&  \frac{-\alpha}{\pi} \int^1_0 d\alpha_1 d\alpha_2 d\alpha_3 \delta\left(1 -\sum \alpha_i \right) \times  \\ \nonumber
&& \left [ \frac{2 m^2 (1-\alpha_2 -\alpha_3) - q^2 (1-\alpha_2 -\alpha_3) - m^2 (\alpha_2 + \alpha_3)^2 - \alpha_2 \alpha_3 q^2}{2 (\alpha_1 m^2_\gamma + (\alpha_2 + \alpha_3)^2 m^2 - \alpha_ 2 \alpha_3 q^2)} +\gamma_{\mbox{{\tiny Euler}}}\right],
\end{eqnarray} 
\begin{eqnarray}
F_2(q^2)&=&  \frac{\alpha}{\pi} \int^1_0 d\alpha_1 d\alpha_2 d\alpha_3 \delta\left(1 -\sum \alpha_i \right) \frac{m^2 \a_1 (\a_2+\a_3)}{\a_1 m^2_\gamma + (\a_2 +\a_3)^2 m^2 -\a_2 \a_3 q^2},
\end{eqnarray} 
\begin{eqnarray}
G_1&=& \frac{\alpha}{\pi} \frac{ i m \gamma_{\mbox{{\tiny Euler}}}}{6}
\end{eqnarray} 
\begin{eqnarray}
E_2&=& 0,
\end{eqnarray} 
\begin{eqnarray}
H_3&=& \frac{-\alpha}{\pi} \frac{3i  \gamma_{\mbox{{\tiny Euler}}}}{4}.
\end{eqnarray} 
The form factors $F_1(q^2)$ and $F_2(q^2)$  are respectively the coefficients of $\gamma^\mu$ and $\frac{i \sigma^{\mu\nu} q_\nu}{2 m}$.

\bigskip
\subsection*{Acknowledgments}
\noindent 
The author is grateful to Jean Iliopoulos and Carmelo Martin for enlightening discussions.
This work was supported in part by the IISN and the Belgian science
policy office (IAP V/27).

\bigskip



\baselineskip=1.6pt
\begin{thebibliography}{99}

\bibitem{Douglas:2001ba}
  M.~R.~Douglas and N.~A.~Nekrasov,
  Rev.\ Mod.\ Phys.\  {\bf 73}, 977 (2001)
  [arXiv:hep-th/0106048].
  
\bibitem{Calmet:2004ii}
  X.~Calmet,
  Phys.\ Rev.\ D {\bf 71}, 085012 (2005)
  [arXiv:hep-th/0411147].



\bibitem{Madore:2000en}
  J.~Madore, S.~Schraml, P.~Schupp and J.~Wess,
  Eur.\ Phys.\ J.\ C {\bf 16}, 161 (2000)
  [arXiv:hep-th/0001203].


\bibitem{Jurco:2000ja}
  B.~Jurco, S.~Schraml, P.~Schupp and J.~Wess,
  Eur.\ Phys.\ J.\ C {\bf 17}, 521 (2000)
  [arXiv:hep-th/0006246].


\bibitem{Jurco:2001rq}
  B.~Jurco, L.~Moller, S.~Schraml, P.~Schupp and J.~Wess,
  Eur.\ Phys.\ J.\ C {\bf 21}, 383 (2001)
  [arXiv:hep-th/0104153].

\bibitem{Calmet:2001na}
  X.~Calmet, B.~Jurco, P.~Schupp, J.~Wess and M.~Wohlgenannt,
  Eur.\ Phys.\ J.\ C {\bf 23}, 363 (2002)
  [arXiv:hep-ph/0111115].

\bibitem{Seiberg:1999vs}
  N.~Seiberg and E.~Witten,
  JHEP {\bf 9909}, 032 (1999)
  [arXiv:hep-th/9908142].


\bibitem{Carlson:2001sw}
  C.~E.~Carlson, C.~D.~Carone and R.~F.~Lebed,
  Phys.\ Lett.\ B {\bf 518}, 201 (2001)
  [arXiv:hep-ph/0107291].

\bibitem{Martin:2002nr}
  C.~P.~Martin,
  Nucl.\ Phys.\ B {\bf 652}, 72 (2003)
  [arXiv:hep-th/0211164].

\bibitem{Wulkenhaar:2001sq}
  R.~Wulkenhaar,
  JHEP {\bf 0203}, 024 (2002)
  [arXiv:hep-th/0112248].


\bibitem{Martin:1999aq}
  C.~P.~Martin and D.~Sanchez-Ruiz,
  Phys.\ Rev.\ Lett.\  {\bf 83}, 476 (1999)
  [arXiv:hep-th/9903077].


\bibitem{Hayakawa:1999yt}
  M.~Hayakawa,
  Phys.\ Lett.\ B {\bf 478}, 394 (2000)
  [arXiv:hep-th/9912094].

\bibitem{Hayakawa:1999zf}
  M.~Hayakawa,
  arXiv:hep-th/9912167.


\bibitem{Riad:2000vy}
  I.~F.~Riad and M.~M.~Sheikh-Jabbari,
  JHEP {\bf 0008}, 045 (2000)
  [arXiv:hep-th/0008132].

\bibitem{Calmet:2004dn}
  X.~Calmet,
  Eur.\ Phys.\ J.\ C {\bf 41}, 269 (2005)
  [arXiv:hep-ph/0401097].

\bibitem{Kersting:2001zz}
  N.~Kersting,
  Phys.\ Lett.\ B {\bf 527}, 115 (2002)
  [arXiv:hep-ph/0109224].
  
  
\bibitem{Calmet:2003jv}
  X.~Calmet and M.~Wohlgenannt,
  Phys.\ Rev.\ D {\bf 68}, 025016 (2003)
  [arXiv:hep-ph/0305027].

\bibitem{Minwalla:1999px}
  S.~Minwalla, M.~Van Raamsdonk and N.~Seiberg,
  JHEP {\bf 0002}, 020 (2000)
  [arXiv:hep-th/9912072].


\end{thebibliography}
\end{document}